\documentclass[runningheads]{llncs}

\usepackage[T1]{fontenc}
\usepackage[utf8]{inputenc}
\usepackage[expansion=false]{microtype}

\usepackage{amsmath}
\usepackage{amssymb}
\usepackage{mathtools}

\usepackage{graphicx}
\graphicspath{{figures/}}
\usepackage{tikz}
\usetikzlibrary{calc,positioning,arrows.meta}
\usepackage{subcaption}
\usepackage{xcolor}
\usepackage{mathrsfs}
\usepackage{placeins}
\usepackage{booktabs,tabularx,array}
\usepackage{xurl}

\usepackage[numbers,sort&compress]{natbib}
\usepackage{hyperref}
\hypersetup{hidelinks}

\newcommand{\KL}[2]{D_{\mathrm{KL}}\!\left[#1\,\middle\|\,#2\right]}
\newcommand{\E}{\mathbb E}

\definecolor{costc}{RGB}{40,80,170}
\definecolor{newc}{RGB}{200,120,30} 

\tikzset{
  bel/.style={circle,draw=black!70,line width=0.4pt,minimum size=7mm,inner sep=0pt,font=\small},
  obs/.style={rectangle,draw=black!45,fill=white,minimum size=2.6mm,inner sep=0pt},
  core/.style={line width=0.9pt,black!75},
  belt/.style={line width=0.5pt,black!55},
  sens/.style={dashed,line width=0.4pt,black!40},
  anno/.style={font=\scriptsize\itshape,text=black!75,align=center},
  annoarrow/.style={-{Stealth[length=1.4mm]},line width=0.3pt,black!55},
  dcore/.style={-{Stealth[length=1.5mm]},line width=0.9pt,black!75},
  dbelt/.style={-{Stealth[length=1.3mm]},line width=0.5pt,black!55},
  ag/.style={circle,draw=black!70,line width=0.5pt,fill=black!6,minimum size=5.2mm,inner sep=0pt,font=\scriptsize},
}

\title{Can Network Models of Science Do Without Pluralist Agents?}
\author{Jonas Hallgren\inst{1,2} \and David Hyland\inst{3} \and Mahault Albarracin\inst{4,5}}
\authorrunning{J. Hallgren et al.}
\institute{Equilibria Network, Stockholm, Sweden
   \and Uppsala University, Uppsala, Sweden
   \and University of Oxford, Oxford, UK
   \and Laboratoire d'Analyse Cognitive de l'Information, Universit\'{e} du Qu\'{e}bec \`{a} Montr\'{e}al, Qu\'{e}bec,
  Canada
   \and Centre of Excellence for AI and Robotics, Sheffield Hallam University, Sheffield, UK}

\begin{document}
\maketitle

\begin{abstract}
Science is a process of useful models being proven wrong time and again, but despite its many flaws, it somehow works. Why is this? Paul Feyerabend argued that it works because people dare to hold and defend incompatible rival theories against the accepted one, since a theory's failures often lead to its abandonment in the presence of a more accurate or simpler alternative. He called this view \textit{theoretical pluralism}. We introduce a theoretical and experimental model in which this can be examined. It compares communities of pluralist agents, which can hold rival theories side by side and maintain beliefs within and between those theories, with communities of monist agents, which can only incorporate what they hear or observe into a single model of their own. In a multi-agent variational inference setting, our experimental demonstrations illustrate how communities with rival theories can go through periods of uncertainty followed by a shift in the theories they hold true, how the information that is communicated between communities affects beliefs when those communities make different observations about the world, and how sharing support for theories and sharing beliefs within them contribute differently to collective belief dynamics.

\end{abstract}

\section{Introduction}
  \label{sec:background}
  \label{sec:hidden}
  If we are to make progress in science, we need to understand the role that diversity of thought plays in it. The question becomes more pressing as Large Language Models (LLMs) are increasingly integrated into the scientific enterprise \citep{messeri2024}. How do we know that they help the scientific process, rather than leading us down locally valid paths whose underlying theoretical assumptions are fundamentally wrong, and what representation of the world would AI systems require to avoid this?

  In this paper, we introduce the idea of epistemically pluralist agents, inspired by theoretical pluralism. This is partly inspired by the work of Paul Feyerabend, who argued that the failures of an accepted theory often show only against an incompatible alternative, so a community that works with one theory at a time hides its own anomalies \citep{feyerabend1975,bschir2015}. One of the practical implications that Feyerabend argued for was that scientists and scientific communities should cultivate and preserve competing theories in order to make sure that we have many ways of viewing and explaining the same phenomena \citep{feyerabend1970}. We can contrast this with its opposite view, \emph{theoretical monism}, where a community works within one theory until incompatible evidence evidence forces it out, as in Kuhn's normal science \citep{kuhn1962}. Which of the two serves a community better has mostly been argued from historical cases \citep{chang2012}, since the question is hard to test directly on real science. This is one reason to turn to models.

  In the last two decades there has been progress on the front of modelling the dynamics of scientific communities~\citep{zollman2007,zollman2013,oconnor2019}. Network-epistemology models study how beliefs about competing options change when agents exchange evidence~\citep{zollman2007,zollman2010,zollman2013,jonard2025}. Opinion-dynamics models also examine communication of numerical estimates~\citep{degroot1974,hegselmann2002}, while active-inference models can represent structured latent states, learning, and belief sharing~\citep{albarracin2022,friston2023federated,catal2024}. Our focus is on the distinction between communicating a conditional belief within a candidate theory and communicating support for that theory relative to its rivals. In order to model this, we require a rich enough representation to describe whether a theory is right or not, and we need to consider the structure of the theories themselves \citep{quine1951,thagard1990,henderson2010}. 
  
  The central distinction we make concerns two levels of belief. On one hand, an agent may be uncertain about the values predicted by a particular theory, and on the other hand, they may be separately uncertain about which theory to favour. We model pluralist agents as those who retain both a conditional belief state for each candidate theory and support between candidates. In contrast, we model monist agents as those who entertain only one belief distribution over observable quantities. One of the central questions we study is what changes when the two levels of belief can be communicated separately.

  What happens if we introduce structure to the models with Bayes nets \citep{pearl1988}? In our model, each agent entertains a set of rival candidate theories, each one a Bayes net over the same observations, together with a belief within each theory and a weight for how much it favours each of them. It can share either the beliefs or the weights with its neighbours. These are our epistemically pluralist agents, and we compare them with monist agents that see the same evidence but only hold one model. We study in our experiments how features such as the observations that agents make and the way that they share information influence the dynamics of individual and collective belief updating. We examine this by looking at a classic philosophy of science example, namely theories of combustion, which has been explored in the philosophy of science literature~\citep{thagard1990,chang2012,kusch2015,blumenthal2017}. When the evidence turns against the accepted theory, pluralist agents go through a period of uncertainty and then switch theory, while monist agents only shift their estimates, with no sign that their explanation has changed. Sharing which theory one favours is what spreads a change of theory, while sharing beliefs within a theory mostly improves predictions. Groups that see different parts of the world can also converge on different theories while still communicating with each other. Going back to the question we started with, our results suggest that an AI system that is to help science, rather than narrow it, should be able to hold more than one theory at a time and communicate beliefs about those theories in addition to beliefs within candidate theories.

\section{Belief Representation and Updating}
\label{sec:bmebmr}

We start by introducing our model in the general setting and then describe experiments using a motivating example based on a competition between the late 18th century theory of phlogiston and the oxygen theory of combustion.

\subsection{Beliefs within and between theories}
\label{sec:carryover}

We model a population of $N$ agents making observations about a set of random variables and communicating synchronously on a fixed network over a series of rounds that occur sequentially through time. In each round, indexed by $t=0,\ldots,T-1$, all agents make an observation at the start of the round, communicate information pertaining to their beliefs within and across theories, evaluate the new observation, and then update their beliefs, with state $t+1$ including observation $t$. Agents are indexed by $i,j\in\{1,\ldots,N\}$, and hold $K$ candidate \textit{theories}, indexed by $k\in\{1,\ldots,K\}$, which represent different probabilistic relationships between the random variables. We use the terms candidate and theory interchangeably in this paper. Theory $k$ represents a vector of represented quantities $x_k\in\mathcal X_k$, where $\mathcal X_k$ is theory $k$'s state space. Some coordinates may be observable and others latent, and different theories may use different latent coordinates. At the beginning of round $t$, agent $i$ maintains two objects for each candidate: a probability $q_{ik,t}$ assigned to the candidate, and a probability density $\rho_{ik,t}(x_k)$ over the quantities represented within that candidate. We parameterize the candidate probabilities by unnormalized log support
$a_{ik,t}\in\mathbb R$:
\begin{equation}
 q_{ik,t}=\frac{\exp \left(a_{ik,t}\right)}{\sum_{k'=1}^K\exp \left(a_{ik',t}\right)},
 \label{eq:state}
\end{equation}
for each theory $k$. Thus, $q_{ik,t}$ represents which theory the agent supports, whereas $\rho_{ik,t}$ represents what values the agent assigns to the quantities represented by theory $k$. The collection $\{q_{ik,t},\rho_{ik,t}\}_{k=1}^K$ constitutes agent $i$'s belief state at round $t$. The real numbers $a_{ik,t}$ store each theory's log weights. Each distribution $\rho_{ik,t}$ has a fixed reference $\rho^0_{ik}$ which serves as a default anchor toward which beliefs relax through forgetting.

\begin{figure}[!htbp]
 \centering\includegraphics[width=\linewidth]{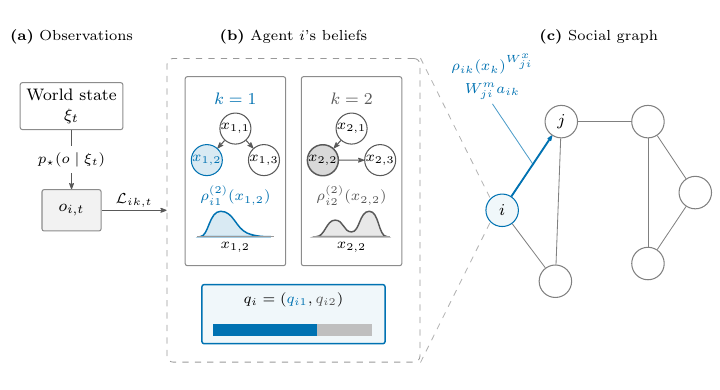}
 \caption{(a) Private observations generated from a common world state whose likelihoods are evaluated under different candidate theories; (b) Agent $i$ retains a conditional belief distribution $\rho_{ik}$ over each candidate's quantities and one model probability $q_{ik}$ for that candidate; (c) Edges on a communication graph represent the exchange of these beliefs.}
 \label{fig:representation}
\end{figure}

\subsection{Environments, observations, and theories}

Let $\mathcal S$ denote the environmental state space and $\xi_t\in\mathcal S$ denote the state of the environment at round $t$. Let $\mathcal O$ denote the observation space. The environment generates observations according to
\begin{equation}
 o_{i,t}\sim p_\star(o\mid \xi_t),
 \label{eq:environment}
\end{equation}
where $p_\star : \mathcal S \to \Delta (\mathcal O)$ denotes the data-generating process.  The environmental state
$\xi_t$ is distinct from the candidate-specific quantity vector $x_k$, in that $\xi_t$ describes the system generating the observations, whereas $x_k$ describes how candidate theory $k$ represents and explains those observations.

Each theory specifies its own likelihood $\mathcal L_{ik,t}(o\mid x_k)$. The theory need not represent the environment in the same coordinates as $\xi_t$, and its likelihood need not coincide with the true data-generating process $p_\star$.

\subsection{Communication, prediction, and belief-updating}
  \label{sec:update}
  We denote the communication matrices for within-theory beliefs and model support by $W^x_t$ and $W^m_t$, respectively. Their entries are nonnegative, each row sums to one, and entry $ij$ represents sender $j$'s influence on receiver $i$ over conditional beliefs and model weights. Quantities with bars are used to denote beliefs after communication within round $t$, and are given by
  \begin{equation}
   \bar\rho_{ik,t}(x_k)\propto\prod_{j=1}^N\rho_{jk,t}(x_k)^{W^x_{ij,t}},\qquad
   \bar a_{ik,t}=(1-\alpha_m)a_{ik,t}
         +\alpha_m\sum_{j=1}^N W^m_{ij,t}a_{jk,t}.
   \label{eq:sharing}
  \end{equation}
  The first operation normalizes the weighted product of beliefs within each candidate, forming a logarithmic opinion pool \citep{genest1986}. The second is a log-linear social-learning rule~\citep{molavi2018}. The parameter $\alpha_m\in[0,1]$ parameterizes model-weight sharing, and the sharing of conditional beliefs uses $W^x_t$ directly. Setting $W^x_t=I_N$ (where $I_n$ is the $n$-dimensional identity matrix) disables conditional sharing, and setting $\alpha_m=0$ disables model-weight sharing.

  \noindent Each agent $i$ evaluates observation $y_{i,t}$ using a normalized likelihood $\mathcal L_{ik,t}(y\mid x_k)$, which can incorporate assigned observation weights. Its predictive log score and subsequent update are given by
  \begin{align}
   \ell_{ik,t}&=\log\int_{\mathcal{X}_k} \mathcal L_{ik,t}(y_{i,t}\mid x_k)
                                  \bar\rho_{ik,t}(x_k)\,dx_k,
   \label{eq:score}\\
   \rho^+_{ik,t}(x_k)&=e^{-\ell_{ik,t}}
         \mathcal L_{ik,t}(y_{i,t}\mid x_k)\bar\rho_{ik,t}(x_k),\qquad
   a^+_{ik,t}=\bar a_{ik,t}+\ell_{ik,t},
   \label{eq:learning}
  \end{align}
  where the superscript $+$ denotes the learned state before reference anchoring is applied. Scoring precedes learning from the same observation, in that a theory can accommodate an observation while receiving a low score for its earlier
  prediction.

  \noindent We model belief updating as a process of minimizing variational free energy over theory distributions
  $\varrho$, which is given by
  \begin{equation}
   \mathcal F_{ik,t}[\varrho]
   =\underbrace{\KL{\varrho}{\bar\rho_{ik,t}}}_{\text{complexity}} -\underbrace{\E_{\varrho}\log \mathcal
  L_{ik,t}(y_{i,t}\mid x_k)}_{\text{accuracy}},
   % =\KL{\varrho}{\rho^+_{ik,t}}-\ell_{ik,t},
   \label{eq:free-energy}
  \end{equation}
  where $\KL{\cdot}{\cdot}$ is the Kullback--Leibler divergence. Thus, we write an agent $i$'s initial belief update
  (before reference anchoring) as
  \begin{equation}
   \rho^+_{ik,t}
   =
   \arg\min_{\varrho}\mathcal F_{ik,t}[\varrho].
   \label{eq:vfe-minimizer}
  \end{equation}

\subsection{Reference anchoring}
We include reference anchoring to represent the persistence of background theoretical commitments. New observations modify an agent’s beliefs, but those beliefs remain partly constrained by the assumptions encoded in the candidate theory’s reference distribution. To model this, we introduce a retention parameter which controls how quickly accumulated evidence can move beliefs away from that reference. Specifically, after seeing the current observation, the agent partially relaxes its within-theory belief toward the fixed reference distribution $\rho_{ik}^0$ for that theory according to
\begin{equation}
 \rho_{ik,t+1}(x_k)\propto
       [\rho^0_{ik}(x_k)]^{1-\omega}[\rho^+_{ik,t}(x_k)]^{\omega},\qquad
 a_{ik,t+1}=a^+_{ik,t},\quad 0\leq\omega\leq1,
 \label{eq:memory}
\end{equation}
where $\omega$ is a retention parameter. At $\omega=1$, the learned belief is retained without relaxation and at $\omega=0$, it is reset to the reference. Because this operation follows the current observation update, $\omega<1$ attenuates both previously accumulated information and the information supplied by the current observation.  Information learned from observation $t$ can affect another agent through communication beginning in round $t+1$.

\section{Experimental setup}
\label{sec:design}
We base our experiments on a simplified model of two theories of combustion from the late 18th century. This motivating example draws on historical debates about the nature of combustion, whose theories and development were more varied than this simplified pair of models~\citep{blumenthal2017,holmes2000,boantza2011}. In particular, the \textit{phlogiston} theory of combustion postulated that flammable objects contain a substance known as phlogiston, which is released upon combustion and absorbed by the surrounding air. According to this theory, substances should lose mass after combustion due to the loss of phlogiston, which made it difficult to accommodate the observation that some substances actually \textit{gained} mass after combustion. This observation was an important factor that contributed to the discovery of oxygen's role in combustion. In our experiments, we use these two theories and associated experimental observations as variables in the networks that agents maintain beliefs over.

\subsection{Theory explanations and observations}
The experiments use multivariate normal (Gaussian) conditional beliefs, described by a mean and covariance. Two different theories share observables for combustion extent $C$, net gas uptake $G$, and solid mass change $M$. The `release' account introduces an unobserved quantity $P$ for hypothetical phlogiston release, and the `oxygen' account introduces unobserved oxygen incorporation $O$. 

\noindent In the experiments, both theory spaces are four-dimensional, i.e., 
$\mathcal X_k=\mathbb R^4$. Their represented state vectors are
\begin{equation*}
 x_{\mathrm P}=(\zeta_{\mathrm P},C,G,M)^\top,
 \qquad
 x_{\mathrm O}=(\zeta_{\mathrm O},C,G,M)^\top .
\end{equation*}

The theory graphs, coefficients, and reference distributions are fixed throughout an experiment. Agents do not learn graph topology or structural coefficients, but only update beliefs about the quantities represented within each theory and update their relative support for each candidate.

The reference observable means are $(1,1,-1)^\top$ for the phlogiston theory and $(1,1,1)^\top$ for the oxygen theory, and the theory-specific latent means are zero. All agents begin with their working beliefs equal to these references and with equal probabilities over the two theories.

In the experiments, the environmental state is specialized to the three common observable quantities,
\[
 \xi_t=(C_t,G_t,M_t)^\top\in\mathbb R^3,
\]
and observations are noisy measurements of this state:
\begin{equation}
 o_{i,t}
 =
 \xi_t+\epsilon_{i,t},
 \qquad
 \epsilon_{i,t}\sim\mathcal N(0,\sigma^2I_3).
 \label{eq:experimental-observation}
\end{equation}

The state $\xi_t$ is constant in the selective-observation comparison and changes during the evidence-shift experiments. Attention weights $w_{i,t}\in[0,1]^3$ specify the precision assigned to each channel. With $H=[\,0_{3\times1}\ I_3\,]$ and $S_{i,t}=\operatorname{diag}(\sqrt{w_{i,t}})$, agents evaluate
\begin{equation}
 y_{i,t}=S_{i,t}o_{i,t},\qquad
 \mathcal L_{ik,t}(y\mid x_k)
 =\mathcal N(y;S_{i,t}Hx_k,\sigma^2I_3).
 \label{eq:attention}
\end{equation}
A positive channel weight $w$ assigns effective channel variance $\sigma^2/w$. These weights change the working likelihood, but not the physical sampling distribution. Appendix~\ref{app:gaussian} gives the Gaussian update.

\subsection{Experimental comparisons}
  \label{sec:comparisons}

  The experiments study how access to evidence, communication, and the representation of alternatives affect belief revision. We distinguish \emph{within-theory sharing}, in which agents exchange beliefs about the quantities represented by each candidate, from \emph{model-support sharing}, in which agents exchange their relative confidences in the competing theories. The former communicates what a theory predicts assuming that it is correct, whereas the latter communicates the evidence accumulated in its favour. We examine these in three comparisons (settings in Appendix~\ref{app:experiments}).

  \paragraph{Revision with full observation.}
  Sixteen agents communicate on a complete graph and initially observe the signal $(C,G,M)=(1,1,-1)$. The mass component then changes sign, giving $(1,1,1)$, and we follow the transition from phlogiston to oxygen (Fig.~\ref{fig:cycle}).

  \paragraph{Partial observation and competing theories.}
  The state is held fixed at $(1,1,1)$. Eight agents observe gas uptake and eight observe mass change, with within-theory sharing at total weight $\beta=0.05$ across groups (Fig.~\ref{fig:matched}). We compare cases where agents keep their model support private against sharing it, as well as a baseline control where all agents maintain a single joint Gaussian belief over combustion, gas uptake, and mass change. In all cases, agents share their conditional beliefs, i.e., their predictions and uncertainties under each theory that they represent. In the two cases where agents share their model supports or keep them private, the agents maintain separate Gaussian beliefs over the observables for each candidate theory.

  \paragraph{Communication channels after changed evidence.}
  We return to the evidence-change scenario from the first experiment on a complete graph, with four fully observing agents and twenty restricted agents that receive only weak gas uptake and mass change evidence (Fig.~\ref{fig:network-channels}). The four conditions share the same history and observation streams and differ only in which communication channels remain enabled after the signal changes. This demonstrates which messages spread support for the oxygen theory and whether the resulting theory choices also improve prediction.

\subsection{Measured outcomes}
\label{sec:outcomes}

We find it useful to distinguish theory choice, numerical disagreement, and predictive accuracy. Theory choice is measured by the probability $q_{i\mathrm O,t}$ weight assigned to the oxygen theory. Group means summarize collective theory support, whereas individual probabilities illustrate how support varies across agents.

Mean individual model entropy describes uncertainty between theories. It is low when agents strongly favour one theory and high when each agent assigns appreciable probability to both theories. It therefore describes a different quantity from disagreement between confident agents.

Numerical disagreement is the distance between the groups' predicted means on the common observable coordinates $(C,G,M)$. For candidate models, these predictions take an average over theories using their current probabilities, and for the Gaussian controls, they use the single Gaussian mean. Different theory choices can yield similar predictions, and similar theory choices can coexist with different numerical estimates.

Predictive accuracy is evaluated over all three observable coordinates, including those receiving little or no weight during learning. In particular, mean squared error (MSE) compares the predicted mean with the generating signal for the evaluated state $\xi_{t-1}$. Negative log-likelihood (NLL) evaluates the complete predictive distribution on independent held-out observations. Lower values indicate better prediction. Appendix~\ref{app:measurements} gives the formulae, reporting windows, and treatment of repeated simulation runs.

\section{Results}
\label{sec:results}

\subsection{Delayed belief revision after changed evidence}
\label{sec:cycle}

With full observation, the community first comes to favour the phlogiston theory under the negative mass signal, then revises toward the oxygen theory after the signal changes. Revision is delayed, in that the new observations must alter the support accumulated during the earlier regime. The model distinguishes this change in preferred explanation from the continuing adjustment of beliefs within each candidate.

\begin{figure}[!ht]
 \centering\includegraphics[width=\linewidth]{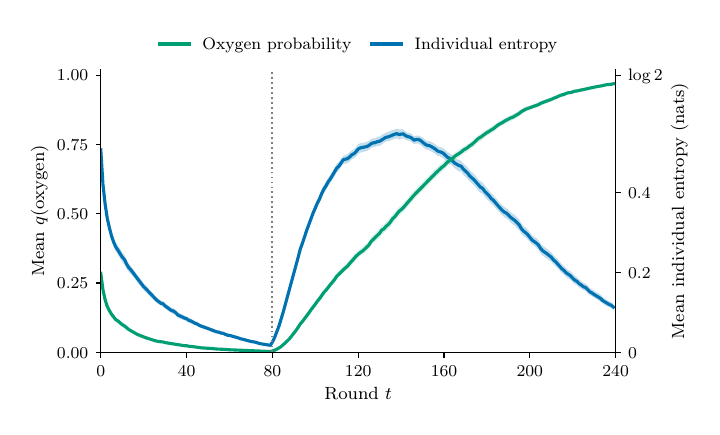}
\vspace{-2em}
 \caption{A demonstration of theory revision under full observation of observable quantities (extent of combustion, net gas uptake, and solid mass change). Mean oxygen probability (green) and mean individual candidate entropy (blue) after the mass signal changes. The dotted line marks the first changed observation at round 80, and the bands represent pointwise 95 percentile bootstrap confidence intervals for means across 50 simulation runs, using 2,000 resamples of whole runs.}
 \label{fig:cycle}
\end{figure}

In this setting, we observe that individual entropy rises while support moves between the theories and falls as agents become confident again. This identifies a period of uncertainty within agents, which is analogous to a period of `crisis' in science. With two theories, such a rise accompanies movement through intermediate candidate theory probabilities.

\subsection{Theory choice and disagreement under selective observation}
\label{sec:separation}
\label{sec:matched}

\begin{figure}[!htb]
 \centering\includegraphics[width=0.94\linewidth]{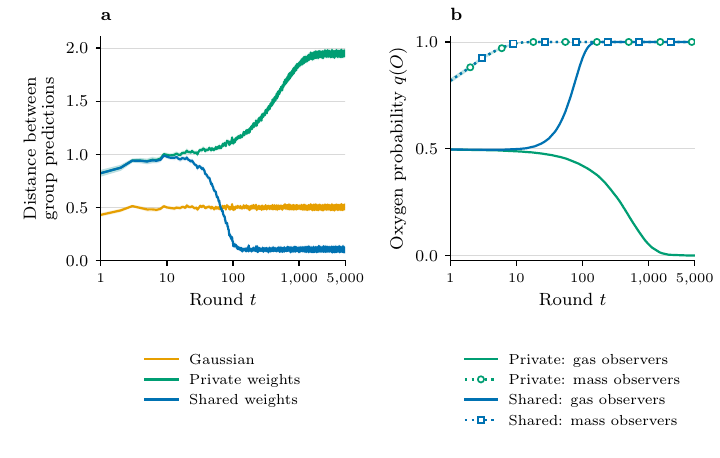}
 \vspace{-1em}
 \caption{Theory choice and numerical prediction under selective observation. All conditions use the same 50 observation streams, $\beta=0.05$, and $\omega=0.3$. (a) Distance between the groups' observable predictive means, where theory predictions are averaged using their current probabilities. (b) Oxygen support in the gas observer group (solid) and mass observer group (dotted), for only the models that maintain beliefs over both theories.}
 \label{fig:matched}
\end{figure}

Different access to the same world can sustain opposing theory preferences even while agents communicate. With within-theory sharing and private model support, gas observers favour the phlogiston theory and mass observers favour the oxygen theory (Fig.~\ref{fig:matched}b). Sharing model support instead leads both groups to favour oxygen. Thus, communicating what each theory predicts does not necessarily align preferences between theories.

The single-Gaussian control highlights what the distance between group predictions measures. This control retains differences between the groups' numerical predictions (Fig.~\ref{fig:matched}a), despite having no probability distribution over alternative theories. This demonstrates that a numerical gap between the two communities' predictions does not by itself establish opposition between theories, and a larger gap does not imply better prediction.

We also observe from Fig.~\ref{fig:matched}b that the support of mass observers for the oxygen theory is virtually unaffected by whether model support is communicated over the network, whereas the beliefs of gas observers in the oxygen theory depend critically on the model support information shared by the mass observers. This demonstrates that sharing predictions and uncertainties alone is not sufficient for the agents observing strong evidence in favour of oxygen to influence the gas-observing agents to favour the oxygen theory.

\subsection{Sharing model support spreads revision after changed evidence}
\label{sec:network}

\begin{figure}[!htb]
\centering\includegraphics[width=\linewidth]{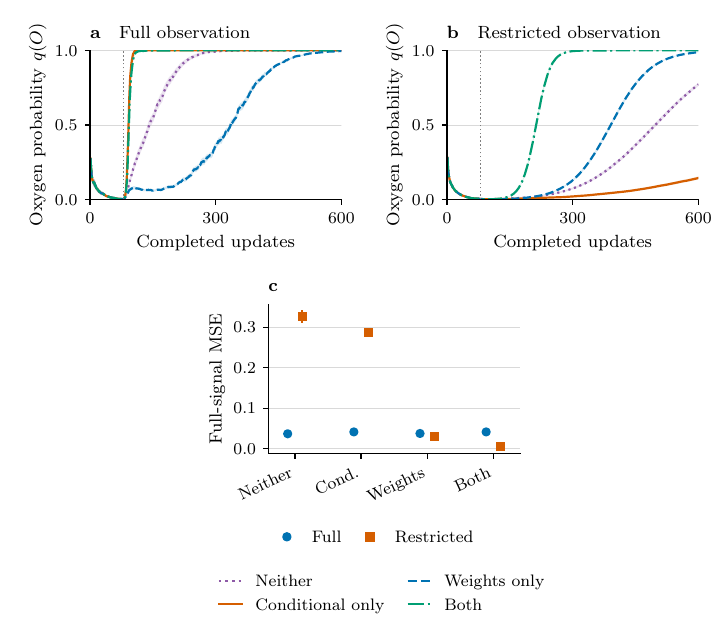}
 \caption{Communication channels after changed evidence on a complete graph. The four conditions begin from the same fully observed, fully communicating state after 80 updates and use the same subsequent observation streams. (a,b) Support for the oxygen theory among four fully observing and twenty restricted agents. The dotted line marks the common state before the first changed observation. (c) Full-signal MSE (i.e., the average squared error in an agent’s predictions across all three observable quantities) is averaged over completed states 451--600, by observation group. Restricted weights are $(1,0.01,0.01)$, $\omega=0.9$, and $\sigma=1$.}
 \label{fig:network-channels}
 \vspace{-2em}
\end{figure}

Having distinguished numerical disagreement from opposing theory preferences under selective observation, we next examine how the two communication channels contribute to revision after a change in evidence. In particular, the final experiment holds the complete graph fixed and changes which messages agents exchange after the signal changes (Fig.~\ref{fig:network-channels}). When model support is shared, either alone or alongside within-theory beliefs, the restricted group's mean oxygen probability approaches 1 over the reported window. Neither condition with private model support reaches that threshold within the simulated horizon, indicating a slower rate of updating in favour of the oxygen theory. Direct communication of accumulated support therefore enables revision in these runs among agents receiving only weak evidence of the change.

Within-theory sharing makes a further contribution on top of model-support sharing. In particular, combined with model-support sharing, it brings the restricted group to believe in the oxygen theory sooner and reduces its late-window MSE relative to sharing model support alone. Communicating model support changes which explanation is favoured, while communicating conditional beliefs also changes the predictions associated with that explanation.

The effects are not uniform across outcomes or observation groups. Within-theory sharing alone modestly improves restricted agents' late-window prediction relative to no communication, while increasing their beliefs in the oxygen theory even more slowly than in the case where no information is shared. Fully observing agents instead have slightly higher late-window MSE when within-theory sharing is enabled. Theory adoption and predictive improvement are therefore distinct outcomes, and the value of a communication channel depends on the evidence available to its recipients.

\section{Discussion}
  \label{sec:discussion}

Can network models of science do without pluralist agents? Our initial results suggest that the answer is no. A monist agent can only share its estimates, whereas a pluralist agent can also share which theory it favours, and these two messages give rise to different belief dynamics. Sharing estimates improved agents' predictions, but agents with weak evidence only switched theory when others told them which theory they favoured. Knowing what a theory predicts does not tell you whether to adopt it, and a model with monist agents cannot see that difference. Pluralist agents can also register when their explanation changes. The local observation update shifts relative support toward the theory assigning the observation the higher predictive probability. A monist agent has no such state, since it can only shift its estimates. This is close to Feyerabend's point that some failures of accepted theories failures become apparent through comparison with alternatives~\citep{feyerabend1975}.

  More experiments are needed to understand under what conditions this remains true and how it interfaces more with existing network epistemological theories. In particular, we have not yet tested how much of a theory's internal structure matters, as opposed to simply holding several theories. The present comparisons do not test the benefit of retaining a currently disfavoured theory, as both theories remain available throughout, and agents do not choose experiments. A natural next step is to let agents choose their experiments and examine how network structure affects discovery and long-term predictive performance when several theories remain available~\citep{zollman2010}.

  Several features also limit the interpretation of these demonstrations. The candidate theories, observation schedules, and attention weights are supplied in advance. Agents do not invent alternatives, revise their graphs, choose which experiments to conduct, or selectively report/update their beliefs based on other motivational considerations~\citep{hyland2025}. These results therefore do not isolate a benefit of explicit latent graphs, or conclusively establish that pluralism improves scientific discovery. Rather, they demonstrate how retaining separate candidates makes theory support distinguishable from numerical adjustment of beliefs.

  The way that agents communicate theories to each other matters, and so does the way that they represent these theories. Telling others which theory you favour changes which theory others favour, as when a few informed agents brought the rest over to the oxygen theory, while telling them what you believe within a theory mostly improves their predictions.

  What this means for science in the age of modern AI systems such as LLMs is an open question, as we do not understand the extent to which they are able to represent and maintain diverse theories. An LLM that folds everything it observes into one account could behave like a monist agent, able to refine a theory but not to notice when a rival explains the data better. An LLM that summarizes what theories predict, without keeping track of which theory the evidence favours, would share estimates but never verdicts. We believe that this needs to be studied and understood, so that science can remain a functioning endeavour with many disagreements, even with automated scientists.

\label{endmain}

\subsection*{Acknowledgements}

The work of D.H. was supported by a UKRI AI World Leading Researcher Fellowship (grant number EP/W002949/1) awarded to Michael Wooldridge.

\vspace{4em}

{\let\clearpage\relax
 \let\cleardoublepage\relax
\bibliographystyle{splncs04}
\bibliography{references}}

\clearpage

\clearpage

\appendix
\renewcommand{\theHsection}{appendix.\Alph{section}}

\section{Mathematical details}
\label{app:math}

We begin by briefly restating the notation and update order of the main text. Communication produces $\bar\rho_{ik,t}$ and $\bar a_{ik,t}$, observation $t$ is scored and learned from, and reference anchoring produces stored state $t+1$. The superscript $+$ denotes the learned state before anchoring. All experiments use positive Gaussian densities and common reference distributions across agents, $\rho^0_{ik}=\rho^0_k$. Communication and anchoring are separate operations from local Bayesian learning.

\subsection{Variational form of within-theory learning}
\label{app:variational}

Writing $Z_{ik,t}=\exp(\ell_{ik,t})$, the predictive normalization and posterior are
\begin{align}
 Z_{ik,t}
 &=\int_{\mathcal{X}_k} \mathcal L_{ik,t}(y_{i,t}\mid x_k)
                  \bar\rho_{ik,t}(x_k)\,\mathrm d x_k,
 \label{eq:app-evidence}\\
 \rho^+_{ik,t}(x_k)
 &=Z_{ik,t}^{-1}\mathcal L_{ik,t}(y_{i,t}\mid x_k)
                       \bar\rho_{ik,t}(x_k).
 \label{eq:app-posterior}
\end{align}
For any normalized trial density $\varrho$ for which the terms are finite,
\begin{align}
 \mathcal F_{ik,t}[\varrho]
 &=\KL{\varrho}{\bar\rho_{ik,t}}
   -\mathbb E_{\varrho}\!\left[\log\mathcal L_{ik,t}(y_{i,t}\mid x_k)\right]
   \notag\\
 &=\KL{\varrho}{\rho^+_{ik,t}}-\log Z_{ik,t}.
 \label{eq:app-vfe}
\end{align}
Thus, $\rho^+_{ik,t}$ minimizes the local objective, with minimum given by $-\ell_{ik,t}$.

\subsection{Reference anchoring and within-theory sharing}
\label{app:anchoring}

The anchoring operation has its own variational form:
\begin{equation}
 \mathcal A_{ik,t}[\varrho]
 =(1-\omega)\KL{\varrho}{\rho_k^0}
        +\omega\KL{\varrho}{\rho^+_{ik,t}}.
 \label{eq:app-anchor-objective}
\end{equation}
For $0\leq\omega\leq1$, its minimizing density is
\begin{equation}
 \rho_{ik,t+1}(x_k)=
 \frac{[\rho_k^0(x_k)]^{1-\omega}[\rho^+_{ik,t}(x_k)]^\omega}
 {\int_{\mathcal{X}_k}[\rho_k^0(v)]^{1-\omega}[\rho^+_{ik,t}(v)]^\omega\,\mathrm d v}.
 \label{eq:app-anchor}
\end{equation}
Substitution allows us to write $\mathcal A_{ik,t}$ as a KL divergence from this density plus a constant. The endpoint $\omega=0$ resets the belief to the reference, whereas $\omega=1$ retains the learned belief. Anchoring applies to within-theory densities, not to the accumulated log support $a_{ik,t}$.

Within-theory communication uses the normalized geometric pool
\begin{equation}
 \bar\rho_{ik,t}(x_k)\propto
            \prod_j\rho_{jk,t}(x_k)^{W^x_{ij,t}}.
 \label{eq:app-within-pool}
\end{equation}
The nonnegative weights sum to one in each row and include self-influence. Only densities under the same candidate are pooled. Agents are assumed to agree on candidate identities and coordinate meanings, and alignment between incompatible vocabularies is not modelled here.

\subsection{Gaussian implementation and selective observation}
\label{app:gaussian}

For $\rho_{ik,t}=\mathcal N(\mu_{ik,t},\Sigma_{ik,t})$, let
\begin{equation}
 \Pi_{ik,t}=\Sigma_{ik,t}^{-1},\qquad h_{ik,t}=\Pi_{ik,t}\mu_{ik,t}.
 \label{eq:app-natural}
\end{equation}
The shared projection is $H=[\,0_{3\times1}\ I_3\,]$ in both theory coordinate systems. Letting $S_{i,t}=\operatorname{diag}(\sqrt{w_{i,t}})$, $y_{i,t}=S_{i,t}o_{i,t}$ and $\widetilde H_{i,t}=S_{i,t}H$, agent $i$'s working likelihood is
\begin{equation}
 \mathcal L_{ik,t}(y\mid x_k)
       =\mathcal N(y;\widetilde H_{i,t}x_k,\sigma^2I_3).
 \label{eq:app-working-likelihood}
\end{equation}
The physical observation noise has covariance $\sigma^2I_3$ before weighting. Consequently, the actual covariance of $S_{i,t}o_{i,t}$ is $\sigma^2S_{i,t}^2$.

\subsection{Model-support dynamics}
\label{app:model-support}

Equations~\eqref{eq:sharing}--\eqref{eq:memory} imply that \begin{equation}
 q_{ik,t+1}\propto
   \exp(\ell_{ik,t})q_{ik,t}^{1-\alpha_m}
                   \prod_jq_{jk,t}^{\alpha_m W^m_{ij,t}},
 \label{eq:app-support-update}
\end{equation}
normalized over candidates. For two candidates, let
\[
z_{i,t}=a_{i\mathrm O,t}-a_{i\mathrm P,t}
       =\log\frac{q_{i\mathrm O,t}}{q_{i\mathrm P,t}},
\qquad s_{i,t}=\ell_{i\mathrm O,t}-\ell_{i\mathrm P,t}.
\]
Then,
\begin{equation}
 z_{i,t+1}=(1-\alpha_m)z_{i,t}
       +\alpha_m\sum_jW^m_{ij,t}z_{j,t}+s_{i,t}.
 \label{eq:odds}
\end{equation}
Within-theory sharing changes the predictive distributions entering $s_{i,t}$ and can therefore affect theory preference even when $\alpha_m=0$. The probabilities express support relative to the supplied candidates under this update rule.

\section{Experimental implementation}
\label{app:experiments}

In this appendix, we provide more details about the implementation of our experiments.\footnote{Accompanying code for our experiments and the reproduction of our figures can be found at \url{https://github.com/dkhyland/epistemic-pluralism}.}

\subsection{Theory reference models}
\label{app:references}

Each reference is compiled from the acyclic linear model
\begin{equation}
 x_k=b_k+B_kx_k+\varepsilon_k,\qquad
 \varepsilon_k\sim\mathcal N(0,R_k).
\end{equation}
In coordinate order $(\zeta_k,C,G,M)$,
\begin{gather}
 \mu_{\mathrm P}^0=(0,1,1,-1)^\top,\qquad
 \mu_{\mathrm O}^0=(0,1,1,1)^\top,\notag\\
 R_k=\operatorname{diag}(0.25,0.5,0.5,0.5).
\end{gather}
Writing $P=\zeta_{\mathrm P}$ and $O=\zeta_{\mathrm O}$, the release model has coefficients $C\to P:+0.8$, $C\to G:+0.8$, and $P\to M:-0.8$; the oxygen model has $C\to O:+0.8$, $O\to G:+0.8$, and $O\to M:+0.8$. All other coefficients are zero. Entry $(B_k)_{vu}$ represents $u\to v$. With $F_k=I-B_k$, we have
\begin{equation}
 b_k=F_k\mu_k^0,\qquad
 \Pi_k^0=F_k^\top R_k^{-1}F_k,\qquad
 h_k^0=F_k^\top R_k^{-1}b_k.
 \label{eq:compile}
\end{equation}
All agents start at their references with equal candidate support and graphs, coefficients, and references remain fixed during each run.

\begin{figure}[!htbp]
 \centering\includegraphics[width=0.82\linewidth]{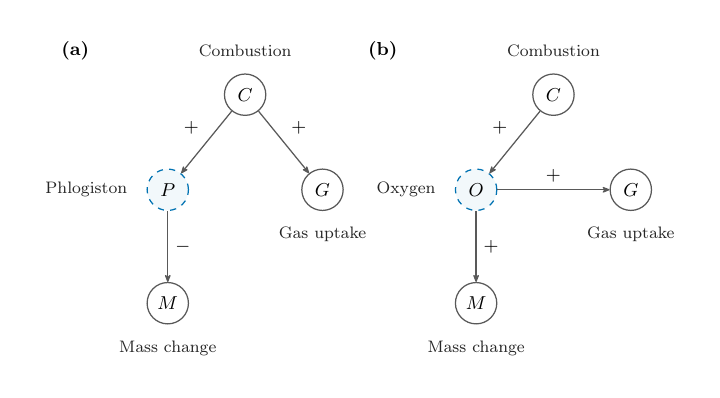}
 \caption{Reference dependencies for the release and oxygen candidates. Dashed nodes are candidate-specific latent quantities; $C$, $G$, and $M$ are shared observables. Arrows give the signs of coefficients of magnitude $0.8$.}
 \label{fig:model}
\end{figure}

The observable reference covariance is $H(\Pi_k^0)^{-1}H^\top$. With common references, initialization at those references, and observable-only likelihoods, all subsequent beliefs factor as $\rho_{ik,t}(\zeta_k,u)=p_k^0(\zeta_k\mid u)\rho^u_{ik,t}(u)$, where $u=(C,G,M)^\top$.

\subsection{Communication between observation groups}
\label{app:communication}

The selective-observation experiment has two groups of $n=8$ agents and constant signal $(C,G,M)=(1,1,1)$. Agents 1--8 use attention weights $(0,1,0)$ and agents 9--16 use $(0,0,1)$; neither group assigns likelihood weight to $C$. For $J_n=\mathbf1_n\mathbf1_n^\top/n$, we have
\begin{equation}
 W_\beta=\begin{pmatrix}
 (1-\beta)J_n&\beta J_n\\ \beta J_n&(1-\beta)J_n
 \end{pmatrix}.
 \label{eq:camp-matrix}
\end{equation}
Each agent assigns total influence $1-\beta$ to its own group, including self-influence, and $\beta$ to the other group. Both candidate conditions use $W^x=W_\beta$. Private support uses $\alpha_m=0$, and shared support uses $W^m=W_\beta$ with $\alpha_m=0.02$. The Gaussian comparison uses the same observations and $W^x=W_\beta$.

\subsection{Single-Gaussian comparison}
\label{app:baselines}

The single-Gaussian comparison represents only $u=(C,G,M)^\top$ and has no candidate theory probabilities. Its common reference is the equally weighted logarithmic pool of the two observable reference marginals. Defining
$\mu_k^{u,0}=H\mu_k^0$ and $\Sigma_k^{u,0}=H(\Pi_k^0)^{-1}H^\top$, we have
\begin{equation}
 \Pi_{\mathrm G}^0=\tfrac12\sum_{k\in\{\mathrm P,\mathrm O\}}
                         (\Sigma_k^{u,0})^{-1},\qquad
 h_{\mathrm G}^0=\tfrac12\sum_{k\in\{\mathrm P,\mathrm O\}}
                         (\Sigma_k^{u,0})^{-1}\mu_k^{u,0}.
 \label{eq:pooled-reference}
\end{equation}
It begins at this reference and uses the same local update, retention, attention weights, and communication matrix as the candidate comparison, with $H=I_3$.

\subsection{Evidence changes and network communication}
\label{app:networks}

All evidence-change experiments use
\begin{equation}
 \xi_t=\begin{cases}(1,1,-1)^\top,&0\leq t<80,\\
                         (1,1,1)^\top,&t\geq80,
             \end{cases}
\end{equation}
with $\sigma=1$ and $\omega=0.9$. The full-observation demonstration (Fig.~\ref{fig:cycle}) has 16 agents, $W_{ij}=1/16$, 240 updates, attention weights $(1,1,1)$ throughout, and both communication channels are enabled with $\alpha_m=0.02$.

The communication and topology comparisons have 24 agents and 600 updates. All agents use $(1,1,1)$ for the first 80 observations. From observation 80, agents 1--4 retain full weights while agents 5--24 use $(1,0.01,0.01)$. All enabled model-support channels use $\alpha_m=0.02$. The complete-graph communication comparison branches from the same fully
observing, fully communicating state after 80 updates in each run, and uses identical subsequent observations in its four conditions:
\begin{table}[!htbp]
 \centering\small
 \label{tab:app-channels}
 \begin{tabular}{@{}lcc@{}}\toprule
 Condition & $W^x$ & $\alpha_m$ \\\midrule
 Neither & $I_{24}$ & $0$\\
 Within-theory only & $W$ & $0$\\
 Model support only & $I_{24}$ & $0.02$\\
 Both & $W$ & $0.02$\\\bottomrule
 \end{tabular}
 \vspace{0.5em}
 \caption{Communication settings from observation 80 in Fig.~\ref{fig:network-channels}.
 Local observation updates remain enabled in every condition. Here $W_{ij}=1/24$.}
\end{table}

\subsection{Revision across communication networks}
\label{sec:topology}

We conducted an additional experiment comparing different network topologies, which keeps both channels enabled, holding each network fixed throughout its initial and changed-evidence periods. We number agents 1--24, so the set of fully-observing agents is $\{1,2,3,4\}$. The ring topology joins consecutive vertices and 24 to 1. The wheel has hub 1, joined to all vertices 2--24, with a cycle through those rim vertices. The community graph has cliques on 1--4 and 5--24 and bridges $(1,8),(2,8),(2,20),(3,9),(3,11),(3,23),(4,24)$. The complete graph joins every distinct pair.

\begin{figure}[!htbp]
 \centering\includegraphics[width=\linewidth]{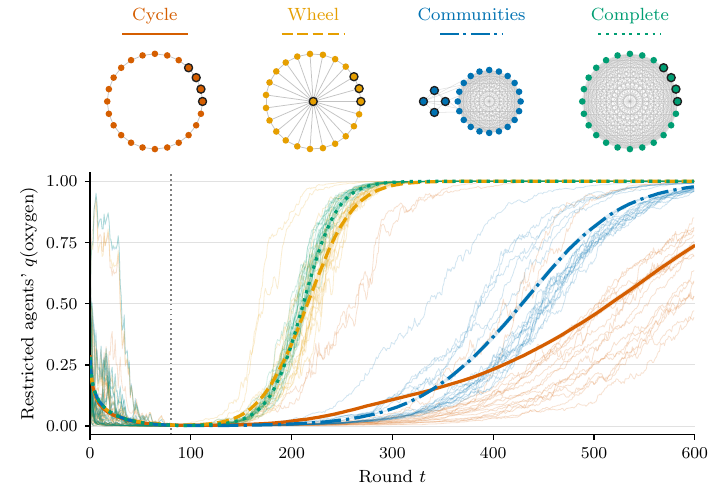}
 \caption{Belief revision of the restricted agents over candidate theories with both channels enabled across four networks. Black outlines around nodes in the topology depictions mark the fully observing agents. Bold curves average the restricted-group mean oxygen support across 50 runs. Bands represent pointwise 95\% percentile-bootstrap intervals from 2,000 whole-run resamples. Thin curves show individual restricted agents in run 0.}
 \label{fig:topology}
\end{figure}

These results show that revision spreads more quickly on the complete network and hub-informed wheel and more slowly on the community network. Restricted-group oxygen support on the cycle network continues to increase at the end of the horizon.

\section{Measurements and reporting}
\label{app:measurements}

\subsection{Timing convention}
\label{app:timing}

Stored state $t$ enters round $t$, and observation $t$ produces state $t+1$. The selective and communication-channel plots label completed states, whereas the cycle and topology plots label generating observation indices and show the corresponding post-update states. Predictive evaluation at state $t\geq1$ uses signal $\xi_{t-1}$, the signal that generated the most recently learned observation.

\subsection{Theory choice and uncertainty}
\label{app:theory-measures}

For group $\mathcal I$, define
\begin{equation}
 \bar q_{\mathcal I,t}(\mathrm O)=
         |\mathcal I|^{-1}\sum_{i\in\mathcal I}q_{i\mathrm O,t}.
\end{equation}
Individual uncertainty is given by
\begin{equation}
 \mathcal H(q_{i,t})=-\sum_kq_{ik,t}\log q_{ik,t},
 \label{eq:app-entropy}
\end{equation}
with $0\log0=0$. The plotted mean averages individual
entropies over agents before averaging runs. It is different from the entropy of a group's mean probability, in that confident disagreement can coexist with low individual entropy.

\subsection{Prediction and numerical disagreement}
\label{app:prediction}

At stored state $t$, a candidate model predicts
\begin{align}
 \hat u_{i,t}&=\sum_kq_{ik,t}H\mu_{ik,t},
 \label{eq:prediction}\\
 p_{i,t}(o)&=\sum_kq_{ik,t}\mathcal N
      (o;H\mu_{ik,t},H\Sigma_{ik,t}H^\top+\sigma^2I_3).
 \label{eq:app-full-prediction}
\end{align}
Evaluation uses all three observable coordinates irrespective of attention during learning. For the Gaussian control, these are the corresponding single-Gaussian mean and predictive density with observation noise added.
For $t\geq1$,
\begin{equation}
 \mathrm{MSE}_{i,t}=\tfrac13\|\hat u_{i,t}-\xi_{t-1}\|_2^2.
\end{equation}
MSE measures estimation error relative to the noiseless signal.

For gas and mass groups $\mathcal I_G$ and $\mathcal I_M$, the distance between community beliefs is given by
\begin{equation}
 D_t=\left\||\mathcal I_G|^{-1}\sum_{i\in\mathcal I_G}\hat u_{i,t}
       -|\mathcal I_M|^{-1}\sum_{i\in\mathcal I_M}\hat u_{i,t}\right\|_2.
 \label{eq:distance}
\end{equation}

\end{document}